\documentclass[aps,prl,amsmath,amssymb,reprint,showkeys,superscriptaddress, %endfloats
]{revtex4-1}

\usepackage{graphicx} % Include figure files
\usepackage{dcolumn}  % Align table columns on decimal point
\usepackage{bm}       % bold math
\usepackage{color}     

\begin{document}

\title{Understanding nanoscale temperature gradients in magnetic nanocontacts}

\author{S. Petit-Watelot}
\author{R. M. Otxoa}
\affiliation{Institut d'Electronique Fondamentale, Universit\'{e} Paris-Sud,
91405 Orsay , France}
\affiliation{UMR 8622, CNRS, 91405 Orsay, France}
\author{M. Manfrini}
\author{W. Van Roy}
\affiliation{IMEC, Kapeldreef 75, B-3001 Leuven, Belgium}
\author{L. Lagae}
\affiliation{IMEC, Kapeldreef 75, B-3001 Leuven, Belgium}
\affiliation{Laboratorium voor Vaste-Stoffysica en Magnetisme, K. U. Leuven, Celestijnenlaan 200 D, B-3001 Leuven, Belgium}
\author{J-V. Kim}
\author{T. Devolder}
\affiliation{Institut d'Electronique Fondamentale, Universit\'{e} Paris-Sud,
91405 Orsay , France}
\affiliation{UMR 8622, CNRS, 91405 Orsay, France}

\keywords{Thermal activation, Vortex, Magnetic dynamics, Point-Contact}

\date{\today}

\begin{abstract}
We determine the temperature profile in magnetic nanocontacts submitted to the very large current densities that are commonly used for spin-torque oscillator behavior.
Experimentally, the quadratic current-induced increase of the resistance through Joule heating is independent of the applied temperature from 6 K to 300 K. The modeling of the experimental rate of the current-induced nucleation of a vortex under the nanocontact, assuming a thermally-activated process, is consistent with a local temperature increase between $150$ K and $220$ K. Simulations of heat generation and diffusion for the actual tridimensional geometry were conducted. They indicate a temperature-independent efficiency of the heat sinking from the electrodes, combined with a localized heating source arising from a nanocontact resistance that is also essentially temperature-independent. For practical currents, we conclude that the local increase of temperature is typically $160$ K and it extends $450$ nm about the nanocontact. Our findings imply that taking into account the current-induced heating at the nanoscale is essential for the understanding of magnetization dynamics in nanocontact systems.
\end{abstract}

\maketitle

%%%%%%%%%%%%%%%%%%%%%%%%%%%%%%%%%%%%%

The discovery of the spin transfer torque \cite{Slonczewski1996, Berger1996} (STT) has provided the ability to manipulate the magnetization with an electrical current. This opened opportunities for new spintronic devices such as spin-torque operated magnetic random access memories or nanosized spin-torque oscillators \cite{Rippard2004}. Unfortunately, STT requires huge current densities, which can lead to substantial heating and early material fatigue. STT may also assist the magnetization switching in nanopillar geometries \cite{Devolder:JAP:2005,Papusoi2008}, and it can increase the magnetization thermal noise \cite{Petit2007}, degrading the device performances. In nanopillar geometries, the magnetization dynamics takes place in a very confined region where the temperature is almost uniform \cite{Lee2008}, such that the temperature dynamics could be understood from simple experiments \cite{Papusoi2008}. In contrast, the temperature rise and its spatial profile remain almost \cite{Devolder:IEEETransMag:2011} unexplored in nanocontact (NC) devices. This is problematic because magnetization dynamics takes place in a much wider area \cite{Mistral2008,Berkov2009} where the temperature might be very non uniform,  potentially affecting the dynamics \cite{Goennenwein2012,Eltschka2010}. In this context, it is essential to develop reliable tools to access to the local temperature during operation under large current densities. 

In this paper, we propose a methodology to access to the local temperature below the NC. The experimental bases of our method rely on the measurement and modeling of the electrical resistance of the nanocontact, combined with the study of the current-pulse-induced vortex nucleation \cite{Devolder2010} at different temperatures. The experimental results are compared with numerical results from tridimensional simulations of the heat generation and diffusion. The simulations allow us to access to the temperature profile. We conclude that at the very large current densities that are commonly used \cite{Rippard2004,Mistral2008}for spin-torque oscillator behavior (around $4~10^{8}$ A/cm$^2$), the temperature increase is typically between $150$ and $220$ K with a very strong spatial gradient, which implies that it is essential to take into account the heating when aiming at understanding magnetization dynamics in nanocontact systems.
    
%%%%%%%%%%%%%%%%%%%%%%%%%%%%%%%%%%%%%%%%% 

\begin{figure}[ht]
	\centering
		\includegraphics[bb=0 0 241 132]{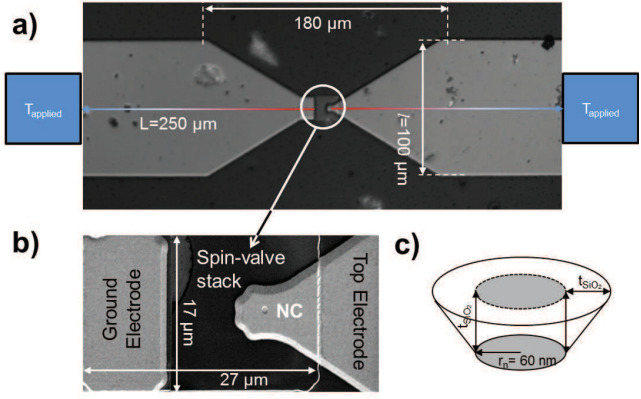}
	\caption{(Color online) (a) Top view of the sample. The bright areas are the gold electrodes. The substrate is covered by 50 nm of SiO$_2$ and appears in black. The sample is contacted with electrical probes on the squares located at the outer edges of the image. At these positions and at the back side of the substrate the temperature is $T_{applied}$. The (colored) arrows show the direction of the heat flux away from the NC. (b) Scanning electron micrograph of the central part of the device, where we can see the $17 \times 27~\mu\textrm{m}^2$ spin valve mesa. As the gold electrodes are deposited in a conformal way, the footprint of the buried NC appears as a circular dot. (c) Sketch of the NC.}
	\label{fig:figure1}
\end{figure}

We work on the NC depicted in Fig. \ref{fig:figure1}. The top electrode forming the nanocontact is made of a Au(200 nm)/Ti(10 nm) truncated cone with a radius at the base $r_{n}=60~nm$ imprinted in a $50$ nm thick SiO$_{2}$ insulating layer (Fig \ref{fig:figure1}c). The NC contacts the top of an extended spin valve (SV) of composition : seed layer 50/IrMn 6/Co$_{90}$Fe$_{10}$ 4.5/Cu 3.5/Ni$_{80}$Fe$_{20}$ 5/Pt 3 (thicknesses are in nm). The stack is grown on a $500~\mu$m thick substrate of intrinsic GaAs. The ends of each electrode (squares in Fig. \ref{fig:figure1}a) are $250~\mu$m away from the NC, and they are electrically contacted with RF probes.

The device magneto-resistance is typically $20~\textrm{m}\Omega$ between the parallel and antiparallel state for a total resistance in the parallel state of $6.2~\Omega$ at room temperature \cite{Kampen2009}. Note that this value includes the resistance of the electrodes.

Low temperature measurements ($5-300$ K) have been performed in vacuum inside a dark cryostat entirely surrounded by a radiative screen. The sample substrate is pressed on a copper thermal chuck, that is maintained at the temperature $T_{applied}$. The electrical probes and the radiative screen are also at $T_{applied}$. To measure the probability of nucleating a vortex state using current pulses, we have used the protocol described in ref. \onlinecite{Devolder2010}: at a given applied temperature, we apply current pulses with amplitudes $43 < I < 48$ mA and durations $\tau_{pulse}$=5-10 ns and measure the resulting microwave voltage spectrum to determine whether a vortex has been created\cite{Devolder2010}. The system is then reset to uniform magnetization, and the procedure is repeated 1000 times, with a waiting time of $100$ ms between each repetition, to get the vortex nucleation probability as a function of $I$, $\tau_{pulse}$ and $T_{applied}$.  

%%%%%%%%%%%%%%%%%%%%%%%%%%%%%%%%%%%%%%%%% 

Let us now summarize our main experimental findings. 

\begin{figure}[htb]
	\centering
		\includegraphics[bb=12 11 253 120]{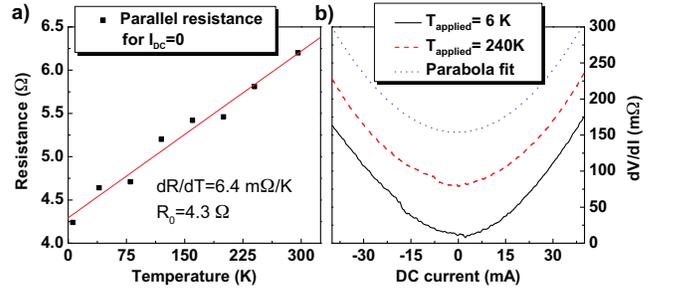}
	\caption{a) Square symbols: resistance of the parallel state for different applied temperatures. Line : linear fit. b) Evolution of the differential resistance with applied DC current for two different applied temperatures, $6$ K (solid line) and $240$ K (dashed line). The upper (dotted) curve is a parabola, used for comparison purpose only. The three curves are vertically offset for clarity.}
	\label{fig:figure2}
\end{figure}

In a first step, we investigate the temperature dependence of the DC electrical properties (Fig. \ref{fig:figure2}). The zero bias resistance of the NC increases linearly with the temperature ($\frac{dR}{d T}=6.4$ m$\Omega$/K) from a residual resistance $R_{0}$ of $4.3~\Omega$. While applying some finite bias current, the differential resistance in the magnetic parallel state increases (fig. \ref{fig:figure2}b) almost with the square of the DC current for all investigated applied temperatures. This indicates that the dominant effect leading to the extra resistance $\delta R$ is Joule heating. The curvature of $\delta R$ is independent of the applied temperature. Surprisingly, this suggests that the current-induced temperature increase is independent of the resistance variations with the temperature in this system.
Our understanding of this unexpected behavior is the following. Let us artificially split the electrical system into two different parts: the Au electrodes on one side, and the nano-contact zone and the spin valve stack (NC+SV) on the other side. We consider that the Au parts are thick enough to be considered as bulk materials, with negligible residual resistance $R_{0_{electrodes}}\approx 0$. The total residual resistance $R_{0}$ is thus essentially that of the NC+SV part, written  $R_{NC+SV}$.
We thus write the total resistance as:

\begin{equation}
\label{ResistanceT}
	R(T)= R_{0} +\left[\left. \frac{dR}{dT} \right| _{NC}+\left. \frac{dR}{dT} \right| _{SV} + \left. \frac{dR}{dT} \right| _{electrodes} \right] T
\end{equation}
	
We believe that electron scattering due to alloy disorder \cite{Counil2006} in the spin-valve and interface reflexions are predominant in $R_{NC+SV}$, which has two consequences. (i) The resistance below the NC is essentially independent of the local temperature (in practice, $\left. \frac{dR}{dT} \right| _{NC}T~\textrm{and}~\left. \frac{dR}{dT}\right| _{SV}T << R_{0}$). (ii) The residual resistance $R_{0}$ is mainly concentrated near the NC/SV contact. The latter point was confirmed \cite{Otxoa2011} by studying the dependence of the total resistance at room temperature with the NC radius, which yielded an effective resistance area product ($RA$) of the NC of $20~\textrm{m}\Omega.\mu m^{2}$ at 300 K. This corresponds to an interface resistance $R_{NC}=2~\Omega$ for the $60$ nm radius NC system. As a consequence, the major part of the Joule losses occurs at the interface between the SV and the NC and it yields a thermal power $\frac{RA}{\pi r_{n}^{2}}I^{2}$, which is independent of the applied temperature. 

This should result in a temperature peak underneath the NC. The heat extraction is ensured by thermal conductivity through the gold top electrode, the rest of the SV stack in series with the gold ground electrode, and the GaAs substrate. Note that the thermal conductivity of bulk gold is almost constant in the temperature window, such that we consider: $\frac{d\kappa_{electrode}}{dT}\approx 0$. We also consider the thermal conductivity of the SV stack and substrate\cite{Ivanov2002} as temperature-independent. Assuming that we can use an effective thermal conductivity $\kappa_{eff}$ relevant for our geometry and our material combination, the temperature increase below the NC ($\delta T_{NC}$) is then independent of $T_{applied}$, and can be written as
\begin{eqnarray}
\label{deltaTNC}
	\delta T_{NC}(I)=\frac{1}{\lambda_{eff} \kappa_{eff}}\frac{RA}{\pi r_{n}^{2}}I^{2}
\end{eqnarray}
where $\lambda_{eff}$ is an effective distance relevant for our geometry. Since the system presents a complex tridimensional geometry, simple considerations are irrelevant to determine quantitatively $\lambda_{eff}$. 

%%%%%%%%%%%%%%%%%%%%%%%%%%%%%%%%%%%%%%%%% 

Instead, we now perform finite element simulations in order to evaluate the temperature rise and its spatial profile. Our solver (Comsol 4.2a) inputs the geometry of Fig. \ref{fig:figure1} and accounts for the electron transport \cite{Petit-Watelot2012} and the resulting Joule heating in the electrode, in the SV stack and in the NC with parameters listed in Table \ref{tab:Tab1}. An interfacial electrical resistance is inserted between the NC and the SV, in line with our previous findings. The Joule-effect-related heat sources are the distributed resistive losses, plus this interfacial resistance. The heat diffuses in the entire volume, including the GaAs substrate and the insulating SiO$_{2}$ layer that surrounds the NC. 

The temperature profile in the stationary regime is calculated with the following boundary conditions : the gold leads terminate at the position of the contact probes, set at $T_{applied}$. The substrate back side is also pinned at $T_{applied}$. The remaining physical boundaries have free temperature conditions, with zero outgoing heat flux.

\begin{table}
	\centering
		\begin{tabular}{|l| c| c|}
			\hline
			Material & Electrical conductivity & Thermal conductivity \\
							 & ($10^{6}$ S/m) & (W/(m.K)) \\
			\hline
			\hline
			Gold bulk & 45 at 296 K & 317 at 296 K \\
								& see Ref. \onlinecite{Handbook} & see Ref. \onlinecite{Handbook} \\
			\hline
			SV stack & 5 & 30 \\
			\hline
			SiO$_{2}$ & 0 & 1.4 \\
			\hline
			Intrinsic GaAs & 0 & 30 \\
			\hline
		\end{tabular}
	\caption{Electrical and thermal conductivities used for the thermal simulations. The parameters of the substrate, SiO$_{2}$ insulating layer and SV are taken as temperature-independent. The SV is considered as a uniform material with average conductivities.}
	\label{tab:Tab1}
\end{table}

\begin{figure}[tb]
	\centering
		\includegraphics[bb=14 11 254 116]{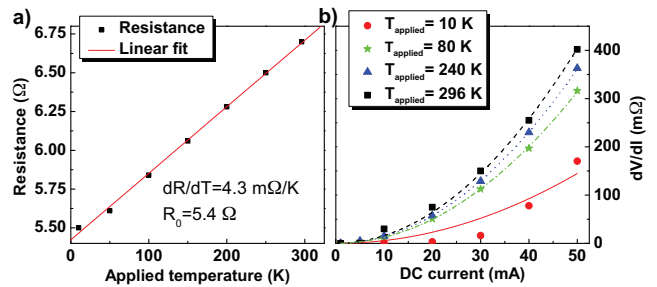}
		\caption {(a) Resistance versus temperature (simulation result). (b) Symbols: (Color online) differential resistance versus DC current (I) for different applied temperatures between 10 K and 296 K (simulation result). Lines: parabolic fits.}
	\label{fig:figure3}
\end{figure}

A first test of the simulation accuracy is to look at the temperature dependence of the resistance of the whole structure (Fig. \ref{fig:figure3}a). The experimentally obtained linear increase of the resistance at warming (Fig. 2a) is reasonably reproduced: the residual resistance $R_{0}=5.4~\Omega$ is $1.1~\Omega$ too large, and the simulated $dR/dT=4.3$ m$\Omega/$K slope underestimates the reality by 33\%. Most of this discrepancy results from our neglecting the temperature dependence of the resistivity of the SV. 

A second test of the simulation accuracy is to look at the current-induced increase of the resistance. This is reported in Fig. \ref{fig:figure3}b for representative applied temperatures between $10~K$ and $296~K$. The simulated $R(I)$ curves are parabolic with a curvature almost independent of the applied temperature, recalling the experimental behavior (Fig. \ref{fig:figure2}b), excepted near 10 K. We find also a parabolic increase of temperature due to Joule heating, almost independent of the applied temperature, of the order of $170$ K for a DC applied current of $48$ mA at the center of the NC/SV interface.

\begin{figure}[tb]
	\centering
		\includegraphics[bb=0 0 242 103]{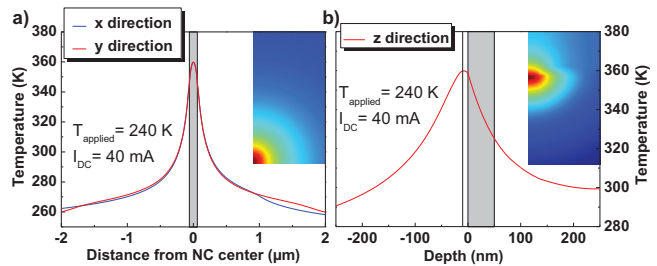}
	\caption{(Color online) Temperature profile, (a) in the plane of the free magnetic layer and (b) in the depth of the sample. Grey areas corresponds respectively to the region below the NC for (a) and of the NC for (b). Fig. a inset : temperature map in the (x,y) plane in a 800 nm  by 400 nm area. The bottom left corner corresponds to the NC center. Fig. b inset : temperature map in the transverse plane (x,z) in a 800 nm  by 400 nm area. The maximum temperature 360 K, is reached at the depth of the free magnetic layer below the NC center. The color contrast scale is from 280 K (blue) to 360 K (red) in both insets.}
	\label{fig:figure4}
\end{figure}

Let us finally look at the simulated temperature profile, either laterally in the free magnetic layer or across the SV thickness underneath the NC (Fig. \ref{fig:figure3}). The temperature profile essentially preserves a cylindrical symmetry around the NC (Fig. \ref{fig:figure4}a inset). It is peaked 5 nm below the NC (Fig. \ref{fig:figure4}b), i.e. right inside the free layer of the SV.  The in-plane temperature distribution decays inversely with the distance outside of the NC (Fig. \ref{fig:figure4}a), with a width at half maximum $\Delta_{warm}$ that varies from $200$ nm to $1200$ nm when the NC radius increases from 40 nm to 80 nm. If we take an average conductivity $\kappa_{eff}=15$ W/(K.m), then the characteristic heat diffusion length $\lambda_{eff}$ is always greater than 1400 nm. Eq. \ref{deltaTNC} can then be viewed as a rule of thumb giving an upper bound for the temperature rise $\delta T_{NC}$ in the NC.

These conclusions are based on electrical measurements that do not inform directly on the temperature in the region of interest for magnetization dynamics. In order to double check our conclusion, we shall now use a different methodology to deduce the local temperature underneath the NC by the study of a thermally activated magnetization process occurring at that precise place, when a current is applied. 

%%%%%%%%%%%%%%%%%%%%%%%%%%%%%%%%%%%%%%%%% 

Our probe of choice is the nucleation of a dynamical vortex state \cite{Pufall2007,Mistral2008,Devolder:APL:2009} as induced by pulsed currents. It has been shown previously \cite{Devolder2010} that the vortex nucleation probability can be described by an Arrh\'{e}nius law with a single activation energy $E_{a}(I)$ which depends only on the total pulse amplitude. Here we shall measure the Arrh\'{e}nius rate to get the real local temperature during the nucleation attempt.
We use the classical description \cite{Lau2006} based on an Arrh\'{e}nius-N\'{e}el law \cite{Neel1949}, for which the mean nucleation time $\left\langle \tau_{nucleation}\right\rangle$ is :
\begin{eqnarray}
	\left\langle \tau_{nucleation}\right\rangle=\tau_{0} \exp{\frac{E_{a}(I)}{k_{B}T}}
\end{eqnarray}
with $T$ the local temperature and $1/\tau_{0}$ the attempt rate. 
%The probability $p_{1}$ of a successful event for one attempt is $p_{1}=\tau_{0}/\left\langle \tau_{nucleation}\right\rangle=e^{-\frac{E_{a}(I)}{k_{B}T}}$. 
The probability $p$ of successful nucleation during the pulse ($\tau_{pulse}$) follows:
\begin{eqnarray}
\label{Nucleation_Probability}
	\ln{(1-p)}=\frac{\tau_{pulse}}{\tau_{0}}\ln{\left[1-\exp{(-\frac{E_{a}}{k_{B}T}})\right]}
\end{eqnarray}
For each parameter set ($I$, $\tau_{pulse}$ and $T_{applied}$), we would in principle need to determine the activation energy $E_{a}(I)$, the attempt rate $\tau_{0}$  and the real local temperature $T=T_{applied} + \delta T_{NC}(I))$. To reduce the number of free parameters, we make the following assumptions. The thermal equilibrium is reached in a time scale shorter than the pulse duration, %the activation energy is independent of time, 
the attempt rate is independent of $I$ and $T$, and the temperature rise ($\delta T_{NC}(I)$) depends only on the pulse amplitude $I$ and not on $T_{applied}$, in line with our previous conclusions. We then use Eq. \ref{Nucleation_Probability} to fit of the evolution of $p$ with the applied temperature. 

\begin{figure}[htb]
	\centering
		\includegraphics[bb=15 14 254 206]{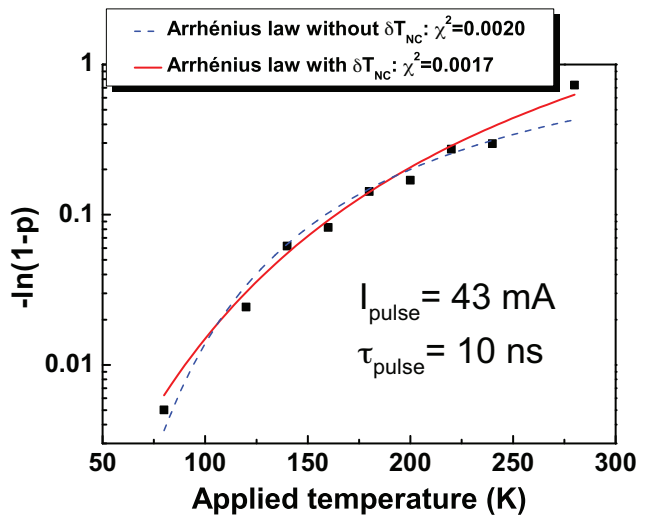}
	\caption{Variation of $-ln(1-p)$ with the applied temperature. Blue dashed line (color only): fit by an Arrhenius law without Joule heating. Red Line (color only): fit by an Arrhenius law taking in to account Joule heating and with a constant $\tau_{0}=250$ ps.}
	\label{fig:Figure5}
\end{figure}
Figure \ref{fig:Figure5} represents the evolution of the nucleation probability with $T_{applied}$ varying from $80$ K to $296$ K for $43$ mA total current and a pulse duration of $10$ ns.  

Two fitting procedures were implemented, and they are compared in figure \ref{fig:Figure5}. (i) In a first step we investigate whether the data can be fitted disregarding the Joule heating (Fig. \ref{fig:Figure5} dash (blue) curve ), i.e. considering $T=T_{applied}$. In that case the free parameters are $E_{a}$ and $\tau_0$. The fitting procedure used is based on the Levenberg-Marquardt method \cite{NumRec} with no weighting of $p$. The results are summarized in Table \ref{Tab:Tab2} for the different parameter sets ($\tau_{pulse}$,$I$) with the normalized residual of the fit. A mean activation time $\tau_0$ of 0.9 ns is found. Importantly, the heating-free fit cannot really reproduce the experimental behavior at low temperature, most probably because the Joule heating induces a substantial error of absolute temperature in that range. 
\begin{table}[tbh]
	\centering
		\begin{tabular}{|r|r|c|c|c|}
			\hline
				$\tau_{pulse}$ & $I_{tot}$ \vline	&	$E_{a}$	in eV without $\delta T_{NC}$ & $\tau_{0}$ in ns & reduced $\chi^{2}$ \\
				\hline
				\hline
				$10$ & $43$ \vline	&	$85$	&	$0.45$	&	$2.0~.10^{-3}$	\\
				\hline
				$10$ & $44$ \vline	&	$62$	&	$0.91$	&	$8.4~.10^{-3}$	\\
				\hline
				$10$ & $48$ \vline	&	$20$	&	$1$	&	$0.2987$	\\
				\hline
				$5$ & $43$ \vline	&	$103$	&	$0.5$	&	$90.0~10^{-6}$	\\
				\hline
				$5$ & $44$ \vline	&	$75$	&	$0.9$	&	$40.0~10^{-6}$	\\
				\hline
				$5$ & $48$ \vline	&	$17$	&	$2$	&	$13.0~10^{-3}$	\\
				\hline
		\end{tabular}
	\caption{Energy barriers $E_{a}(I)$ and inverse attempt frequency $\tau_{0}$ resulting from fits performed on the experimental data disregarding the Joule heating (i.e. $T=T_{applied}$).}
	\label{Tab:Tab2}
\end{table}
(ii) In a second step, we account for the Joule heating. We consider the real temperature as $T=T_{applied}+\delta T_{NC}(I)$, and perform the fitting procedure with fixed $\tau_{0}$ at $250$ ps and with only two free parameters $E_{a}$ and $\delta T_{NC}$. This procedure is examplified in Fig.\ref{fig:Figure5} (red line). The results are listed in table \ref{Tab:Tab3}. The convergence of the fit is significantly improved for low temperatures and for high currents (i.e. low $E_{a}(I)$), i.e. in the cases where the role of the heating is the most important. 
\begin{table}[tbh]
	\centering
		\begin{tabular}{|r|r|c|c|c|}
			\hline
				$\tau_{pulse}$ & $I_{tot}$ & $E_{a}$ in eV & $\delta T_{NC}$ in K & reduced $\chi^{2}$\\
				\hline
				\hline
				$10$ & $43$ \vline	&	$112$	&	$40$ &	$1.73~.10^{-3}$ \\
				\hline
				$10$ & $44$ \vline	&	$121$	&	$90$	& $6.88~.10^{-3}$ \\
				\hline
				$10$ & $48$ \vline	&	$73$	&	$147$	& $0.160$ \\
				\hline
				$5$ & $43$ \vline	&	$136$	&	$39$	& $80.0~.10^{-5}$ \\
				\hline
				$5$ & $44$ \vline	&	$131$	&	$71$	& $20.0~.10^{-5}$ \\
				\hline
				$5$ & $48$ \vline	&	$108$	&	$225$	& $3.93~.10^{-3}$ \\
				\hline
		\end{tabular}
	\caption{Energy barriers $E_{a}(I)$ and current-induced temperature rise resulting from fits performed on the experimental data taking into account Joule heating with fixed $\tau_{0}=0.25$ ns.}
	\label{Tab:Tab3}
\end{table}

From these fits, the rise in temperature due to Joule heating can be written with a satisfactory agreement as $\delta T_{NC}(I)=\eta I^2$ with $\eta \approx 10^{5}$ K/A$^{2}$. For instance, for a total current of $48$ mA, the real temperature increase during the pulse lies between $150$ K and $220$ K, which is consistent with the value of 170 K that we predicted previously using thermal simulations.

%%%%%%%%%%%%%%%%%%%%%%%%%%%%%%%%%%%%%%%%% 

In summary, we have studied the temperature rise and its profile in magnetic nanocontacts submitted to the very large current densities that are commonly used for spin-torque oscillator behavior. We accessed the temperature changes occurring during the application of the current using two experimental methods. They are: the modeling of the transport properties and, the modeling of the nucleation rate of vortex structures. Both methods were implemented in various current conditions and applied temperature environments. We confronted our data to finite element simulations of the current and heat distribution in a realistic tridimensional geometry. For the current densities typically used on nanocontacts in the oscillator regime, the local increase of temperature reaches typically 150 to 220 K and it extends over an area with a typical diameter of $450$ nm about a nanocontact of $60$ nm radius. 

In conclusion, thermal noise may be thus substantially higher than commonly expected, with detrimental consequences when using the nanocontact as spin torque oscillators.  In addition, this heating probably leads to substantial changes in the magnetic properties, for instance in the exchange bias field of the reference layers. Finally, the temperature gradients are unusually strong, and may lead to additional sources of spin torques \cite{Padran-Hernandez2011} that can play a role in the magnetization dynamics. Our findings imply that taking into account the current-induced heating at the nanoscale is essential for the understanding of magnetization dynamics in nanocontact systems.

This work has been supported by the Agence Nationale de la Recherche contract VOICE (ANR-09-NANO-006-01) and the European Community under the 7th FP for the Marie Curie ITN SEMISPINNET, contract no. 215368-2.

%%%%%%%%%%%%%%%%%%%%%%%%%%%%%%%%%%%%%%%%%%%

%bibliography


%%%%%%%%%%%%%%%%%%%%%%%%%%%%%%%%%%%%%%%%%%%
 

\begin{thebibliography}{25}%
\makeatletter
\providecommand \@ifxundefined [1]{%
 \@ifx{#1\undefined}
}%
\providecommand \@ifnum [1]{%
 \ifnum #1\expandafter \@firstoftwo
 \else \expandafter \@secondoftwo
 \fi
}%
\providecommand \@ifx [1]{%
 \ifx #1\expandafter \@firstoftwo
 \else \expandafter \@secondoftwo
 \fi
}%
\providecommand \natexlab [1]{#1}%
\providecommand \enquote  [1]{``#1''}%
\providecommand \bibnamefont  [1]{#1}%
\providecommand \bibfnamefont [1]{#1}%
\providecommand \citenamefont [1]{#1}%
\providecommand \href@noop [0]{\@secondoftwo}%
\providecommand \href [0]{\begingroup \@sanitize@url \@href}%
\providecommand \@href[1]{\@@startlink{#1}\@@href}%
\providecommand \@@href[1]{\endgroup#1\@@endlink}%
\providecommand \@sanitize@url [0]{\catcode `\\12\catcode `\$12\catcode
  `\&12\catcode `\#12\catcode `\^12\catcode `\_12\catcode `\%12\relax}%
\providecommand \@@startlink[1]{}%
\providecommand \@@endlink[0]{}%
\providecommand \url  [0]{\begingroup\@sanitize@url \@url }%
\providecommand \@url [1]{\endgroup\@href {#1}{\urlprefix }}%
\providecommand \urlprefix  [0]{URL }%
\providecommand \Eprint [0]{\href }%
\providecommand \doibase [0]{http://dx.doi.org/}%
\providecommand \selectlanguage [0]{\@gobble}%
\providecommand \bibinfo  [0]{\@secondoftwo}%
\providecommand \bibfield  [0]{\@secondoftwo}%
\providecommand \translation [1]{[#1]}%
\providecommand \BibitemOpen [0]{}%
\providecommand \bibitemStop [0]{}%
\providecommand \bibitemNoStop [0]{.\EOS\space}%
\providecommand \EOS [0]{\spacefactor3000\relax}%
\providecommand \BibitemShut  [1]{\csname bibitem#1\endcsname}%
\let\auto@bib@innerbib\@empty
%</preamble>
\bibitem [{\citenamefont {Slonczewski}(1996)}]{Slonczewski1996}%
  \BibitemOpen
  \bibfield  {author} {\bibinfo {author} {\bibfnamefont {J.~C.}\ \bibnamefont
  {Slonczewski}},\ }\href
  {http://www.sciencedirect.com/science/article/B6TJJ-403WKFH-1/2/fd06569c67005bde75d947ee6cfc273e}
  {\bibfield  {journal} {\bibinfo  {journal} {Journal of Magnetism and Magnetic
  Materials}\ }\textbf {\bibinfo {volume} {159}},\ \bibinfo {pages} {L1}
  (\bibinfo {year} {1996})}\BibitemShut {NoStop}%
\bibitem [{\citenamefont {Berger}(1996)}]{Berger1996}%
  \BibitemOpen
  \bibfield  {author} {\bibinfo {author} {\bibfnamefont {L.}~\bibnamefont
  {Berger}},\ }\href {http://link.aps.org/doi/10.1103/PhysRevB.54.9353}
  {\bibfield  {journal} {\bibinfo  {journal} {Phys. Rev. B}\ }\textbf {\bibinfo
  {volume} {54}},\ \bibinfo {pages} {9353} (\bibinfo {year}
  {1996})}\BibitemShut {NoStop}%
\bibitem [{\citenamefont {Rippard}\ \emph {et~al.}(2004)\citenamefont
  {Rippard}, \citenamefont {Pufall}, \citenamefont {Kaka}, \citenamefont
  {Russek},\ and\ \citenamefont {Silva}}]{Rippard2004}%
  \BibitemOpen
  \bibfield  {author} {\bibinfo {author} {\bibfnamefont {W.~H.}\ \bibnamefont
  {Rippard}}, \bibinfo {author} {\bibfnamefont {M.~R.}\ \bibnamefont {Pufall}},
  \bibinfo {author} {\bibfnamefont {S.}~\bibnamefont {Kaka}}, \bibinfo {author}
  {\bibfnamefont {S.~E.}\ \bibnamefont {Russek}}, \ and\ \bibinfo {author}
  {\bibfnamefont {T.~J.}\ \bibnamefont {Silva}},\ }\href
  {http://link.aps.org/doi/10.1103/PhysRevLett.92.027201} {\bibfield  {journal}
  {\bibinfo  {journal} {Phys. Rev. Lett.}\ }\textbf {\bibinfo {volume} {92}},\
  \bibinfo {pages} {027201} (\bibinfo {year} {2004})}\BibitemShut {NoStop}%
\bibitem [{\citenamefont {Devolder}\ \emph {et~al.}(2005)\citenamefont
  {Devolder}, \citenamefont {Tulapurkar}, \citenamefont {Suzuki}, \citenamefont
  {Chappert}, \citenamefont {Crozat},\ and\ \citenamefont
  {Yagami}}]{Devolder:JAP:2005}%
  \BibitemOpen
  \bibfield  {author} {\bibinfo {author} {\bibfnamefont {T.}~\bibnamefont
  {Devolder}}, \bibinfo {author} {\bibfnamefont {A.}~\bibnamefont
  {Tulapurkar}}, \bibinfo {author} {\bibfnamefont {Y.}~\bibnamefont {Suzuki}},
  \bibinfo {author} {\bibfnamefont {C.}~\bibnamefont {Chappert}}, \bibinfo
  {author} {\bibfnamefont {P.}~\bibnamefont {Crozat}}, \ and\ \bibinfo {author}
  {\bibfnamefont {K.}~\bibnamefont {Yagami}},\ }\href {\doibase
  10.1063/1.2012512} {\bibfield  {journal} {\bibinfo  {journal} {Journal of
  Applied Physics}\ }\textbf {\bibinfo {volume} {98}},\ \bibinfo {eid} {053904}
  (\bibinfo {year} {2005})}\BibitemShut {NoStop}%
\bibitem [{\citenamefont {Papusoi}\ \emph {et~al.}(2008)\citenamefont
  {Papusoi}, \citenamefont {Sousa}, \citenamefont {Herault}, \citenamefont
  {Prejbeanu},\ and\ \citenamefont {Dieny}}]{Papusoi2008}%
  \BibitemOpen
  \bibfield  {author} {\bibinfo {author} {\bibfnamefont {C.}~\bibnamefont
  {Papusoi}}, \bibinfo {author} {\bibfnamefont {R.}~\bibnamefont {Sousa}},
  \bibinfo {author} {\bibfnamefont {J.}~\bibnamefont {Herault}}, \bibinfo
  {author} {\bibfnamefont {I.~L.}\ \bibnamefont {Prejbeanu}}, \ and\ \bibinfo
  {author} {\bibfnamefont {B.}~\bibnamefont {Dieny}},\ }\href
  {http://stacks.iop.org/1367-2630/10/i=10/a=103006} {\enquote {\bibinfo
  {title} {Probing fast heating in magnetic tunnel junction structures with
  exchange bias},}\ } (\bibinfo {year} {2008})\BibitemShut {NoStop}%
\bibitem [{\citenamefont {Petit}\ \emph {et~al.}(2007)\citenamefont {Petit},
  \citenamefont {Baraduc}, \citenamefont {Thirion}, \citenamefont {Ebels},
  \citenamefont {Liu}, \citenamefont {Li}, \citenamefont {Wang},\ and\
  \citenamefont {Dieny}}]{Petit2007}%
  \BibitemOpen
  \bibfield  {author} {\bibinfo {author} {\bibfnamefont {S.}~\bibnamefont
  {Petit}}, \bibinfo {author} {\bibfnamefont {C.}~\bibnamefont {Baraduc}},
  \bibinfo {author} {\bibfnamefont {C.}~\bibnamefont {Thirion}}, \bibinfo
  {author} {\bibfnamefont {U.}~\bibnamefont {Ebels}}, \bibinfo {author}
  {\bibfnamefont {Y.}~\bibnamefont {Liu}}, \bibinfo {author} {\bibfnamefont
  {M.}~\bibnamefont {Li}}, \bibinfo {author} {\bibfnamefont {P.}~\bibnamefont
  {Wang}}, \ and\ \bibinfo {author} {\bibfnamefont {B.}~\bibnamefont {Dieny}},\
  }\href {http://link.aps.org/doi/10.1103/PhysRevLett.98.077203} {\bibfield
  {journal} {\bibinfo  {journal} {Phys. Rev. Lett.}\ }\textbf {\bibinfo
  {volume} {98}},\ \bibinfo {pages} {077203} (\bibinfo {year}
  {2007})}\BibitemShut {NoStop}%
\bibitem [{\citenamefont {Lee}\ and\ \citenamefont {Lim}(2008)}]{Lee2008}%
  \BibitemOpen
  \bibfield  {author} {\bibinfo {author} {\bibfnamefont {D.~H.}\ \bibnamefont
  {Lee}}\ and\ \bibinfo {author} {\bibfnamefont {S.~H.}\ \bibnamefont {Lim}},\
  }\href {http://dx.doi.org/10.1063/1.2943151} {\bibfield  {journal} {\bibinfo
  {journal} {Appl. Phys. Lett.}\ }\textbf {\bibinfo {volume} {92}},\ \bibinfo
  {pages} {233502} (\bibinfo {year} {2008})}\BibitemShut {NoStop}%
\bibitem [{\citenamefont {Devolder}\ \emph {et~al.}(2011)\citenamefont
  {Devolder}, \citenamefont {Kim}, \citenamefont {Petit-Watelot}, \citenamefont
  {Otxoa}, \citenamefont {Chappert}, \citenamefont {Manfrini}, \citenamefont
  {Van~Roy},\ and\ \citenamefont {Lagae}}]{Devolder:IEEETransMag:2011}%
  \BibitemOpen
  \bibfield  {author} {\bibinfo {author} {\bibfnamefont {T.}~\bibnamefont
  {Devolder}}, \bibinfo {author} {\bibfnamefont {J.-V.}\ \bibnamefont {Kim}},
  \bibinfo {author} {\bibfnamefont {S.}~\bibnamefont {Petit-Watelot}}, \bibinfo
  {author} {\bibfnamefont {R.}~\bibnamefont {Otxoa}}, \bibinfo {author}
  {\bibfnamefont {C.}~\bibnamefont {Chappert}}, \bibinfo {author}
  {\bibfnamefont {M.}~\bibnamefont {Manfrini}}, \bibinfo {author}
  {\bibfnamefont {W.}~\bibnamefont {Van~Roy}}, \ and\ \bibinfo {author}
  {\bibfnamefont {L.}~\bibnamefont {Lagae}},\ }\href {\doibase
  10.1109/TMAG.2010.2101056} {\bibfield  {journal} {\bibinfo  {journal}
  {Magnetics, IEEE Transactions on}\ }\textbf {\bibinfo {volume} {47}},\
  \bibinfo {pages} {1595 } (\bibinfo {year} {2011})}\BibitemShut {NoStop}%
\bibitem [{\citenamefont {Mistral}\ \emph {et~al.}(2008)\citenamefont
  {Mistral}, \citenamefont {van Kampen}, \citenamefont {Hrkac}, \citenamefont
  {Kim}, \citenamefont {Devolder}, \citenamefont {Crozat}, \citenamefont
  {Chappert}, \citenamefont {Lagae},\ and\ \citenamefont
  {Schrefl}}]{Mistral2008}%
  \BibitemOpen
  \bibfield  {author} {\bibinfo {author} {\bibfnamefont {Q.}~\bibnamefont
  {Mistral}}, \bibinfo {author} {\bibfnamefont {M.}~\bibnamefont {van Kampen}},
  \bibinfo {author} {\bibfnamefont {G.}~\bibnamefont {Hrkac}}, \bibinfo
  {author} {\bibfnamefont {J.-V.}\ \bibnamefont {Kim}}, \bibinfo {author}
  {\bibfnamefont {T.}~\bibnamefont {Devolder}}, \bibinfo {author}
  {\bibfnamefont {P.}~\bibnamefont {Crozat}}, \bibinfo {author} {\bibfnamefont
  {C.}~\bibnamefont {Chappert}}, \bibinfo {author} {\bibfnamefont
  {L.}~\bibnamefont {Lagae}}, \ and\ \bibinfo {author} {\bibfnamefont
  {T.}~\bibnamefont {Schrefl}},\ }\href
  {http://link.aps.org/doi/10.1103/PhysRevLett.100.257201} {\bibfield
  {journal} {\bibinfo  {journal} {Phys. Rev. Lett.}\ }\textbf {\bibinfo
  {volume} {100}},\ \bibinfo {pages} {257201} (\bibinfo {year}
  {2008})}\BibitemShut {NoStop}%
\bibitem [{\citenamefont {Berkov}\ and\ \citenamefont
  {Gorn}(2009)}]{Berkov2009}%
  \BibitemOpen
  \bibfield  {author} {\bibinfo {author} {\bibfnamefont {D.~V.}\ \bibnamefont
  {Berkov}}\ and\ \bibinfo {author} {\bibfnamefont {N.~L.}\ \bibnamefont
  {Gorn}},\ }\href {http://link.aps.org/doi/10.1103/PhysRevB.80.064409}
  {\bibfield  {journal} {\bibinfo  {journal} {Phys. Rev. B}\ }\textbf {\bibinfo
  {volume} {80}},\ \bibinfo {pages} {064409} (\bibinfo {year}
  {2009})}\BibitemShut {NoStop}%
\bibitem [{\citenamefont {Goennenwein}\ and\ \citenamefont
  {Bauer}(2012)}]{Goennenwein2012}%
  \BibitemOpen
  \bibfield  {author} {\bibinfo {author} {\bibfnamefont {S.~T.~B.}\
  \bibnamefont {Goennenwein}}\ and\ \bibinfo {author} {\bibfnamefont
  {G.~E.~W.}\ \bibnamefont {Bauer}},\ }\href@noop {} {\bibfield  {journal}
  {\bibinfo  {journal} {Nature Nanotechnology}\ }\textbf {\bibinfo {volume}
  {7}},\ \bibinfo {pages} {145} (\bibinfo {year} {2012})}\BibitemShut {NoStop}%
\bibitem [{\citenamefont {Eltschka}\ \emph {et~al.}(2010)\citenamefont
  {Eltschka}, \citenamefont {Wotzel}, \citenamefont {Rhensius}, \citenamefont
  {Krzyk}, \citenamefont {Nowak}, \citenamefont {Klaui}, \citenamefont
  {Kasama}, \citenamefont {Dunin-Borkowski}, \citenamefont {Heyderman},
  \citenamefont {van Driel},\ and\ \citenamefont {Duine}}]{Eltschka2010}%
  \BibitemOpen
  \bibfield  {author} {\bibinfo {author} {\bibfnamefont {M.}~\bibnamefont
  {Eltschka}}, \bibinfo {author} {\bibfnamefont {M.}~\bibnamefont {Wotzel}},
  \bibinfo {author} {\bibfnamefont {J.}~\bibnamefont {Rhensius}}, \bibinfo
  {author} {\bibfnamefont {S.}~\bibnamefont {Krzyk}}, \bibinfo {author}
  {\bibfnamefont {U.}~\bibnamefont {Nowak}}, \bibinfo {author} {\bibfnamefont
  {M.}~\bibnamefont {Klaui}}, \bibinfo {author} {\bibfnamefont
  {T.}~\bibnamefont {Kasama}}, \bibinfo {author} {\bibfnamefont {R.~E.}\
  \bibnamefont {Dunin-Borkowski}}, \bibinfo {author} {\bibfnamefont {L.~J.}\
  \bibnamefont {Heyderman}}, \bibinfo {author} {\bibfnamefont {H.~J.}\
  \bibnamefont {van Driel}}, \ and\ \bibinfo {author} {\bibfnamefont {R.~A.}\
  \bibnamefont {Duine}},\ }\href {\doibase 10.1103/PhysRevLett.105.056601}
  {\bibfield  {journal} {\bibinfo  {journal} {Physical Review Letters}\
  }\textbf {\bibinfo {volume} {105}},\ \bibinfo {pages} {056601} (\bibinfo
  {year} {2010})}\BibitemShut {NoStop}%
\bibitem [{\citenamefont {Devolder}\ \emph {et~al.}(2010)\citenamefont
  {Devolder}, \citenamefont {Kim}, \citenamefont {Manfrini}, \citenamefont {van
  Roy}, \citenamefont {Lagae},\ and\ \citenamefont {Chappert}}]{Devolder2010}%
  \BibitemOpen
  \bibfield  {author} {\bibinfo {author} {\bibfnamefont {T.}~\bibnamefont
  {Devolder}}, \bibinfo {author} {\bibfnamefont {J.-V.}\ \bibnamefont {Kim}},
  \bibinfo {author} {\bibfnamefont {M.}~\bibnamefont {Manfrini}}, \bibinfo
  {author} {\bibfnamefont {W.}~\bibnamefont {van Roy}}, \bibinfo {author}
  {\bibfnamefont {L.}~\bibnamefont {Lagae}}, \ and\ \bibinfo {author}
  {\bibfnamefont {C.}~\bibnamefont {Chappert}},\ }\href
  {http://link.aip.org/link/?APL/97/072512/1} {\bibfield  {journal} {\bibinfo
  {journal} {Appl. Phys. Lett.}\ }\textbf {\bibinfo {volume} {97}},\ \bibinfo
  {pages} {072512} (\bibinfo {year} {2010})}\BibitemShut {NoStop}%
\bibitem [{\citenamefont {van Kampen}\ \emph {et~al.}(2009)\citenamefont {van
  Kampen}, \citenamefont {Lagae}, \citenamefont {Hrkac}, \citenamefont
  {Schrefl}, \citenamefont {Kim}, \citenamefont {Devolder},\ and\ \citenamefont
  {Chappert}}]{Kampen2009}%
  \BibitemOpen
  \bibfield  {author} {\bibinfo {author} {\bibfnamefont {M.}~\bibnamefont {van
  Kampen}}, \bibinfo {author} {\bibfnamefont {L.}~\bibnamefont {Lagae}},
  \bibinfo {author} {\bibfnamefont {G.}~\bibnamefont {Hrkac}}, \bibinfo
  {author} {\bibfnamefont {T.}~\bibnamefont {Schrefl}}, \bibinfo {author}
  {\bibfnamefont {J.-V.}\ \bibnamefont {Kim}}, \bibinfo {author} {\bibfnamefont
  {T.}~\bibnamefont {Devolder}}, \ and\ \bibinfo {author} {\bibfnamefont
  {C.}~\bibnamefont {Chappert}},\ }\href
  {http://stacks.iop.org/0022-3727/42/i=24/a=245001} {\bibfield  {journal}
  {\bibinfo  {journal} {Journal of Physics D: Applied Physics}\ }\textbf
  {\bibinfo {volume} {42}},\ \bibinfo {pages} {245001} (\bibinfo {year}
  {2009})}\BibitemShut {NoStop}%
\bibitem [{\citenamefont {Counil}\ \emph {et~al.}(2006)\citenamefont {Counil},
  \citenamefont {Devolder}, \citenamefont {Kim}, \citenamefont {Crozat},
  \citenamefont {Chappert}, \citenamefont {Zoll},\ and\ \citenamefont
  {Fournel}}]{Counil2006}%
  \BibitemOpen
  \bibfield  {author} {\bibinfo {author} {\bibfnamefont {G.}~\bibnamefont
  {Counil}}, \bibinfo {author} {\bibfnamefont {T.}~\bibnamefont {Devolder}},
  \bibinfo {author} {\bibfnamefont {J.~V.}\ \bibnamefont {Kim}}, \bibinfo
  {author} {\bibfnamefont {P.}~\bibnamefont {Crozat}}, \bibinfo {author}
  {\bibfnamefont {C.}~\bibnamefont {Chappert}}, \bibinfo {author}
  {\bibfnamefont {S.}~\bibnamefont {Zoll}}, \ and\ \bibinfo {author}
  {\bibfnamefont {R.}~\bibnamefont {Fournel}},\ }\href {\doibase
  10.1109/TMAG.2006.879718} {\bibfield  {journal} {\bibinfo  {journal} {Ieee
  Transactions On Magnetics}\ }\textbf {\bibinfo {volume} {42}},\ \bibinfo
  {pages} {3323} (\bibinfo {year} {2006})}\BibitemShut {NoStop}%
\bibitem [{\citenamefont {Otxoa}\ \emph {et~al.}(2011)\citenamefont {Otxoa},
  \citenamefont {Manfrini}, \citenamefont {Devolder}, \citenamefont {Kim},
  \citenamefont {van Roy}, \citenamefont {Lagae},\ and\ \citenamefont
  {Chappert}}]{Otxoa2011}%
  \BibitemOpen
  \bibfield  {author} {\bibinfo {author} {\bibfnamefont {R.~M.}\ \bibnamefont
  {Otxoa}}, \bibinfo {author} {\bibfnamefont {M.}~\bibnamefont {Manfrini}},
  \bibinfo {author} {\bibfnamefont {T.}~\bibnamefont {Devolder}}, \bibinfo
  {author} {\bibfnamefont {J.~V.}\ \bibnamefont {Kim}}, \bibinfo {author}
  {\bibfnamefont {W.}~\bibnamefont {van Roy}}, \bibinfo {author} {\bibfnamefont
  {L.}~\bibnamefont {Lagae}}, \ and\ \bibinfo {author} {\bibfnamefont
  {C.}~\bibnamefont {Chappert}},\ }\href {\doibase 10.1002/pssb.201001208}
  {\bibfield  {journal} {\bibinfo  {journal} {Physica Status Solidi B-basic
  Solid State Physics}\ }\textbf {\bibinfo {volume} {248}},\ \bibinfo {pages}
  {1615} (\bibinfo {year} {2011})}\BibitemShut {NoStop}%
\bibitem [{\citenamefont {Ivanov}\ \emph {et~al.}(2002)\citenamefont {Ivanov},
  \citenamefont {Luk'yanov}, \citenamefont {Merisov}, \citenamefont
  {Sologubenko},\ and\ \citenamefont {Khadjai}}]{Ivanov2002}%
  \BibitemOpen
  \bibfield  {author} {\bibinfo {author} {\bibfnamefont {A.~I.}\ \bibnamefont
  {Ivanov}}, \bibinfo {author} {\bibfnamefont {A.~N.}\ \bibnamefont
  {Luk'yanov}}, \bibinfo {author} {\bibfnamefont {B.~A.}\ \bibnamefont
  {Merisov}}, \bibinfo {author} {\bibfnamefont {A.~V.}\ \bibnamefont
  {Sologubenko}}, \ and\ \bibinfo {author} {\bibfnamefont {G.~Y.}\ \bibnamefont
  {Khadjai}},\ }\href {\doibase 10.1063/1.1491187} {\bibfield  {journal}
  {\bibinfo  {journal} {Low Temperature Physics}\ }\textbf {\bibinfo {volume}
  {28}},\ \bibinfo {pages} {462} (\bibinfo {year} {2002})}\BibitemShut
  {NoStop}%
\bibitem [{\citenamefont {Petit-Watelot}\ \emph {et~al.}(2012)\citenamefont
  {Petit-Watelot}, \citenamefont {Otxoa},\ and\ \citenamefont
  {Manfrini}}]{Petit-Watelot2012}%
  \BibitemOpen
  \bibfield  {author} {\bibinfo {author} {\bibfnamefont {S.}~\bibnamefont
  {Petit-Watelot}}, \bibinfo {author} {\bibfnamefont {R.~M.}\ \bibnamefont
  {Otxoa}}, \ and\ \bibinfo {author} {\bibfnamefont {M.}~\bibnamefont
  {Manfrini}},\ }\href {http://dx.doi.org/10.1063/1.3687915} {\bibfield
  {journal} {\bibinfo  {journal} {Appl. Phys. Lett.}\ }\textbf {\bibinfo
  {volume} {100}},\ \bibinfo {pages} {083507} (\bibinfo {year}
  {2012})}\BibitemShut {NoStop}%
\bibitem [{\citenamefont {Haynes}(2011)}]{Handbook}%
  \BibitemOpen
  \bibinfo {editor} {\bibfnamefont {W.~M.}\ \bibnamefont {Haynes}},\ ed.,\
  \href@noop {} {\emph {\bibinfo {title} {Handbook of Chemistry and
  Physics}}},\ \bibinfo {edition} {92nd}\ ed.\ (\bibinfo  {publisher} {CRC
  Press},\ \bibinfo {year} {2011})\BibitemShut {NoStop}%
\bibitem [{\citenamefont {Pufall}\ \emph {et~al.}(2007)\citenamefont {Pufall},
  \citenamefont {Rippard}, \citenamefont {Schneider},\ and\ \citenamefont
  {Russek}}]{Pufall2007}%
  \BibitemOpen
  \bibfield  {author} {\bibinfo {author} {\bibfnamefont {M.~R.}\ \bibnamefont
  {Pufall}}, \bibinfo {author} {\bibfnamefont {W.~H.}\ \bibnamefont {Rippard}},
  \bibinfo {author} {\bibfnamefont {M.~L.}\ \bibnamefont {Schneider}}, \ and\
  \bibinfo {author} {\bibfnamefont {S.~E.}\ \bibnamefont {Russek}},\ }\href
  {http://link.aps.org/doi/10.1103/PhysRevB.75.140404} {\bibfield  {journal}
  {\bibinfo  {journal} {Phys. Rev. B}\ }\textbf {\bibinfo {volume} {75}},\
  \bibinfo {pages} {140404} (\bibinfo {year} {2007})}\BibitemShut {NoStop}%
\bibitem [{\citenamefont {Devolder}\ \emph {et~al.}(2009)\citenamefont
  {Devolder}, \citenamefont {Kim}, \citenamefont {Crozat}, \citenamefont
  {Chappert}, \citenamefont {Manfrini}, \citenamefont {van Kampen},
  \citenamefont {Roy}, \citenamefont {Lagae}, \citenamefont {Hrkac},\ and\
  \citenamefont {Schrefl}}]{Devolder:APL:2009}%
  \BibitemOpen
  \bibfield  {author} {\bibinfo {author} {\bibfnamefont {T.}~\bibnamefont
  {Devolder}}, \bibinfo {author} {\bibfnamefont {J.-V.}\ \bibnamefont {Kim}},
  \bibinfo {author} {\bibfnamefont {P.}~\bibnamefont {Crozat}}, \bibinfo
  {author} {\bibfnamefont {C.}~\bibnamefont {Chappert}}, \bibinfo {author}
  {\bibfnamefont {M.}~\bibnamefont {Manfrini}}, \bibinfo {author}
  {\bibfnamefont {M.}~\bibnamefont {van Kampen}}, \bibinfo {author}
  {\bibfnamefont {W.~V.}\ \bibnamefont {Roy}}, \bibinfo {author} {\bibfnamefont
  {L.}~\bibnamefont {Lagae}}, \bibinfo {author} {\bibfnamefont
  {G.}~\bibnamefont {Hrkac}}, \ and\ \bibinfo {author} {\bibfnamefont
  {T.}~\bibnamefont {Schrefl}},\ }\href@noop {} {\bibfield  {journal} {\bibinfo
   {journal} {Appl. Phys. Lett.}\ }\textbf {\bibinfo {volume} {95}},\ \bibinfo
  {eid} {012507} (\bibinfo {year} {2009})}\BibitemShut {NoStop}%
\bibitem [{\citenamefont {Lau}\ \emph {et~al.}(2006)\citenamefont {Lau},
  \citenamefont {Bording}, \citenamefont {Beleggia},\ and\ \citenamefont
  {Zhu}}]{Lau2006}%
  \BibitemOpen
  \bibfield  {author} {\bibinfo {author} {\bibfnamefont {J.~W.}\ \bibnamefont
  {Lau}}, \bibinfo {author} {\bibfnamefont {J.~K.}\ \bibnamefont {Bording}},
  \bibinfo {author} {\bibfnamefont {M.}~\bibnamefont {Beleggia}}, \ and\
  \bibinfo {author} {\bibfnamefont {Y.}~\bibnamefont {Zhu}},\ }\href
  {http://link.aip.org/link/?APL/88/012508/1} {\bibfield  {journal} {\bibinfo
  {journal} {Appl. Phys. Lett.}\ }\textbf {\bibinfo {volume} {88}},\ \bibinfo
  {pages} {012508} (\bibinfo {year} {2006})}\BibitemShut {NoStop}%
\bibitem [{\citenamefont {N\'{e}el}(1949)}]{Neel1949}%
  \BibitemOpen
  \bibfield  {author} {\bibinfo {author} {\bibfnamefont {L.}~\bibnamefont
  {N\'{e}el}},\ }\href@noop {} {\bibfield  {journal} {\bibinfo  {journal} {Ann.
  Geophys.}\ }\textbf {\bibinfo {volume} {5}},\ \bibinfo {pages} {99} (\bibinfo
  {year} {1949})}\BibitemShut {NoStop}%
\bibitem [{\citenamefont {Williiam H.~Press}\ and\ \citenamefont
  {Flannery}(2007)}]{NumRec}%
  \BibitemOpen
  \bibfield  {author} {\bibinfo {author} {\bibfnamefont {W.~H.}\
  \bibnamefont {Press}}, \bibinfo {author} {\bibfnamefont {S.~A.}\ \bibnamefont {Teukolsky}}, \bibinfo {author} {\bibfnamefont {W.~T.}\ \bibnamefont {Vetterling}}, \ and\ \bibinfo {author} {\bibfnamefont {B.~P.}\ \bibnamefont {Flannery}},\
  }\href@noop {} {\emph {\bibinfo {title} {Numerical Recipes, chap. 15}}},\
  \bibinfo {edition} {third}\ ed.\ (\bibinfo  {publisher} {Cambridge
  University Press},\ \bibinfo {address} {32 Avenue
  of the Americas, New York, NY 10013-2473, USA},\ \bibinfo {year} {2007})\ p.\
  \bibinfo {pages} {1256}\BibitemShut {NoStop}%
\bibitem [{\citenamefont {Padr\'{o}n-Hern\'{a}ndez}\ \emph
  {et~al.}(2011)\citenamefont {Padr\'{o}n-Hern\'{a}ndez}, \citenamefont
  {Azevedo},\ and\ \citenamefont {Rezende}}]{Padran-Hernandez2011}%
  \BibitemOpen
  \bibfield  {author} {\bibinfo {author} {\bibfnamefont {E.}~\bibnamefont
  {Padr\'{o}n-Hern\'{a}ndez}}, \bibinfo {author} {\bibfnamefont
  {A.}~\bibnamefont {Azevedo}}, \ and\ \bibinfo {author} {\bibfnamefont
  {S.~M.}\ \bibnamefont {Rezende}},\ }\href
  {http://link.aps.org/doi/10.1103/PhysRevLett.107.197203} {\bibfield
  {journal} {\bibinfo  {journal} {Phys. Rev. Lett.}\ }\textbf {\bibinfo
  {volume} {107}},\ \bibinfo {pages} {197203} (\bibinfo {year}
  {2011})}\BibitemShut {NoStop}%
\end{thebibliography}
\end{document}